\documentclass[12pt]{article}
\usepackage{graphicx,amsmath,amssymb,color,bm}

\begin{document}

\centerline {\Large\textbf {Feature-rich electronic excitations in external}}
\centerline {\Large\textbf {fields of 2D silicene}}


\centerline{Jhao-Ying Wu$^{1,\star}$, Szu-Chao Chen, Godfrey Gumbs$^{2,\dag}$, and Ming-Fa Lin$^{1,\dag\dag}$ }
\centerline{$^{1}$Department of Physics, National Cheng Kung University,
Tainan, Taiwan 701}
\centerline{$^{2}$Department of Physics and Astronomy, Hunter College at the City University of New York,\\ \small }
\centerline{695 Park Avenue, New York, New York 10065, USA}

\vskip0.6 truecm

\noindent

Electronic Coulomb excitations in monolayer silicene are investigated by using the Lindhard dielectric function and a newly developed generalized tight-binding model (G-TBM). G-TBM simultaneously contains the atomic interactions, the spin-orbit coupling, the Coulomb interactions, and the various external fields at an arbitrary chemical potential. We exhibit the calculation results of the electrically tunable magnetoplasmons and the strong magnetic field modulation of plasmon behaviors. The two intriguing phenomena are well explained by determining the dominant transition channels in the dielectric function and through understanding the electron behavior under the multiple interactions (intrinsic and external). A further tunability of the plasmon features is demonstrated with the momentum transfer and the Fermi energy. The methodological strategy could be extended to several other 2D materials like germanene and stanene, and might open a pathway to search a better system in nanoplasmonic applications.



\vskip0.6 truecm

$\mathit{PACS}$: 73.22.Lp, 73.22.Pr

\newpage

\bigskip

\centerline {\textbf {I. INTRODUCTION}}%

\bigskip

Silicon photonics is currently a very active and progressive area of research, as silicon optical circuits have emerged as the replacement technology for copper-based circuits in communication and broadband networks. The demand continues for ever improving communications and computing performance, and this in turn means that photonic circuits are finding ever increasing application areas. This paper provides an important and timely investigation of a "high-priority topic" in the field, covering a particular aspect  of the technology that forms the research area of silicon photonics.

\medskip
\par

Ever since the epitaxial synthesis in 2010 of silicene \cite{nineteen,Vogt:2012,Chen:2012,Liu:2014}, a buckled structure in which the silicon atoms are displaced perpendicular to the basal plane, there has been a widespread  effort by researchers to gain knowledge of its atomic and electronic properties. As an excellent candidate material, silicene is featured with its strong SOC and an electrically tunable band gap. A key effect due to the electric field is that it can open and close the energy band gap which is a desired  functionality for digital electronics applications and is one of several potential applications for silicene as a field-effect transistor operating at
room temperature \cite{NatureNano,NatureNano2}. Studied electronic properties include the quantum spin Hall effect, anomalous hall insulators and single-valley semimetals  \cite{nine}, potential giant magneto-resistance \cite{ten}, superconductivity \cite{chiral}, topologically  protected helical edge states \cite{eleven,RC} and other exotic field-dependent phenomena.

The collective-Coulomb excitations, dominated by the e-e interactions, are important to understand the behavior of electrons in a material. In an intrinsic monolayer silicene, low-frequency plasmons hardly exist, mainly due to the vanishing density of states at the Fermi level. The lack of low-frequency plasmons may be improved by additional dopants or a gate insertion to increase the free-charge density \cite{Wunsch2006,SH2009}, i.e., uplifting or lowering the Fermi level to increase the density of states. Alternatively, the free carriers (with zero Fermi energy) may be generated by increasing the thermally excited electrons and holes in the conduction and valence bands, respectively. The intrinsic band gap in silicene would cause the interplay between the intraband and interband transitions and lead to an undamped plasmon at the low frequency [].

A uniform magnetic field would make cyclotron motion of electrons and form the dispersionless Landau levels (LLs), which may enhance the low density of states largely. Unlike graphene, where the n=0 Landau level is pinned at the zero, the stronger spin-orbit coupling (SOC) in silicene causes the n=0 level to split between $\pm\lambda_{so}$/2 ($\lambda_{so}$ is the amplitude of the effective SOC). In the presence of an electric field, the single-valley Landau levels are no longer spin degenerate and the spin-down and -up states would construct their separate energy gaps. Therefore, it is interesting to investigate how the external fields change the plasmon behaviors.

The present paper is based on the investigation of magnetoplasmons in silicene with the consideration of an electric field and a tunable chemical potential. The calculation is performed by our developed generalized tight-binding model, which simultaneously incorporates all meaningful interactions, including the atomic interaction, the spin-orbit coupling, the Coulomb interaction, and the interaction between the electron and the external fields. This makes our results reliable in a wide range of chemical potentials, the field strengths, and the excitation frequency. The magnetoplasmons can be categorized into two groups: a propagating and a localized mode. The former is strongly driven by the e-e Coulomb interaction, while the latter is mainly dominated by the interaction between the electron and the magnetic field. The electric field is to induce more localized plasmon modes due to the lift of spin and valley degeneracy, a result associated with the buckling geometry structure. On the other hand, the heightened Fermi level would bring about a long-lived propagating plasmon mode. We pay particular attention to the B-dependent plasmon spectrum and observe rich changes in the plasmon features when crossing a critical field strength $B_{c}$. The modulation of the plasmon excitations by externally applied electric and magnetic fields means a possible way to design an active plasmon device in the low-buckled materials.

\bigskip
\bigskip
\centerline {\textbf {II. METHODS}}%
\bigskip
\bigskip

Similar to graphene, silicene consists of a honeycomb lattice of silicon atoms with two sublattices made up of A and B sites. The difference is that silicene has a buckled structure, with the two sublattice planes separated by a distance of $2\ell$ with $\ell=0.23$ ${\AA}$ (Fig. 1). In the tight-binding approximation, the Hamiltonian for silicene in the presence of SOC is given in Refs. \cite{CChen:2012-2,2Ezawa:2012}: We have

\begin{equation}
H=-t\sum_{\langle ij\rangle\alpha}c^{\dag}_{i\alpha}c_{j\alpha}+i\frac{\lambda_{SO}}{3\sqrt{3}}\sum_{\langle\langle ij\rangle\rangle\alpha\beta}v_{ij}c^{\dag}_{i\alpha}\sigma^{z}_{\alpha\beta}c_{j\beta}-i\frac{2}{3}\lambda_{R2}\sum_{\langle\langle ij\rangle\rangle\alpha\beta}u_{ij}c^{\dag}_{i\alpha}(\vec{\sigma}\times \vec{d^{0}_{ij}})^{z}_{\alpha\beta}c_{j\beta}+\ell\sum_{i\alpha}\mu_{i}E_{z}c^{\dag}_{i\alpha}c_{i\alpha},
\end{equation}
Here,  $c^{\dag}_{i\alpha}$ and $c_{i\alpha}$ are creation and destruction operators for an electron with spin polarization $\alpha=\uparrow,\downarrow$ at lattice site $i$, when acting on a chosen wavefunction. The sums are carried out over either nearest-neighbor $\langle i,j\rangle$ or next-nearest-neighbor
$\langle\langle i,j \rangle\rangle$ lattice site pairs, as indicated. The first term (I) takes care of nearest-neighbor hopping with  transfer energy $t=1.6$ eV whereas the third term (II) describes the effective SOC for parameter $\lambda_{SOC}=3.9$ meV and $\vec{\sigma}=(\sigma_{x},\sigma_{y},\sigma_{z})$ is the vector of Pauli spin  matrices. Additionally,  we chose $v_{ij}=\pm 1$ if the next-nearest-neighboring hopping is anticlockwise/clockwise  with respect to the positive $z$ axis. In the third term (III), We include the intrinsic Bychkov-Rashba SOC via  $\lambda_{R2}=0.7$ meV for the next-nearest neighbor hopping and set $u_{i}=\pm 1$ for the A/B  lattice site, respectively, and $\hat{d}_{ij}=\mathbf{d_{ij}}/|d_{ij}|$ with the vector $\mathbf{d_{ij}}$ joining two sites $i$ and $j$ on the same sublattice. The staggered sublattice potential energy produced by the external electric field
is described by the fourth term where $\mu_i=\pm 1$ for the A/B sublattice site and  $\ell$=0.23 ${\AA}$.

The monolayer silicene is assumed to be in an ambient uniform magnetic field $\mathbf{B}=B\hat{z}$. The magnetic flux, the product of the field strength and the hexagonal area, is $\Phi= [3\sqrt{3}b^{2}B/2]/\phi_{0}$ where $\phi_{0} = h/e = 4.1356กั1015\  T/m^{2}$. The vector potential,
which is chosen as $\textbf{A} =  (Bx)\hat{y}$, leads to a new periodicity along the armchair direction. The unit cell is thus enlarged and its dimension is determined by $R_{B}= 1/\Phi$. The enlarged unit cell contains 4$R_{B}$ Si atoms and the Hamiltonian matrix is 8$R_{B}$$\times$8$R_{B}$ including the spin degree. The hopping parameter $t$ acquires the extra position-dependent Peierls phase is given by
\begin{equation}
\varphi_{ij}=\exp  \left ( i\frac{e}{\hbar} \int_{\vec{x}_n}  ^{\vec{x}_m} d\vec{x}^\prime
\cdot \vec{A} (\vec{x}^\prime)\right)  \ .
\end{equation}

The phases of the first three terms of the hopping integrals in Eq. (1) associated with the additional position-dependent Peierls phases are given in order:

\begin{eqnarray}
(i)~~~~\langle B^\alpha_{k}|H|A^\alpha_{j}\rangle&&=\sum_{nn}{1\over
N}\exp{[i{\bf k}\cdot({\bf R}_{A^\alpha_{j}}-{\bf R}_{B^\alpha_{k}})]}
\times\exp{\{i[{2\pi\over \phi_0}\int ^{{\bf R}_{B^\alpha_{k}}}_{{\bf
R}_{A^\alpha_{j}}}{\bf A}\cdot d{\bf r}]\}} \cr
&&=t_{1,j}\delta_{j,k}+\gamma_0s\delta_{j,k+1} \ ,
\end{eqnarray}
with $t_{1,j}=\exp\{i[-k_x{b\over 2}-k_y{\sqrt{3}b\over
2}+\pi{\Phi\over \phi_0}(j-1+{1\over 6})]\}$ and $s=\exp[i(-k_{x}b)]$.

\begin{eqnarray}
(ii)~~~~\langle A^\alpha_{k}|H|A^\alpha_{j}\rangle&&=\sum_{nn}{1\over
N}\exp{[i{\bf k}\cdot({\bf R}_{A^\alpha_{j}}-{\bf R}_{A^\alpha_{k}})]}
\times\exp{\{i[{2\pi\over \phi_0}\int ^{{\bf R}_{A^\alpha_{k}}}_{{\bf
R}_{A^\alpha_{j}}}{\bf A}\cdot d{\bf r}]\}} \cr
&&=t_{2,j}\delta_{j,k} \ ,
\end{eqnarray}
where $t_{2,j}=\exp i[k_{y}a+2\pi{\Phi\over\phi_0}(j-1)]-\exp i[-k_{y}a-2\pi{\Phi\over\phi_0}(j-1)]$.

\begin{eqnarray}
\langle B^\alpha_{k}|H|B^\alpha_{j}\rangle&&=\sum_{nn}{1\over
N}\exp{[i{\bf k}\cdot({\bf R}_{B^\alpha_{j}}-{\bf R}_{B^\alpha_{k}})]}
\times\exp{\{i[{2\pi\over \phi_0}\int ^{{\bf R}_{B^\alpha_{k}}}_{{\bf
R}_{B^\alpha_{j}}}{\bf A}\cdot d{\bf r}]\}} \cr
&&=t_{3,j}\delta_{j,k} \ ,
\end{eqnarray}
and $t_{3,j}=\exp i\{-k_{y}a-2\pi{\Phi\over\phi_0}[(j-1)+\frac{1}{3}]\}-\exp i\{k_{y}a+2\pi{\Phi\over\phi_0}[(j-1)+\frac{1}{3}]\}$.

\begin{eqnarray}
\langle A^\alpha_{k}|H|A^\alpha_{j}\rangle&&=\sum_{nn}{1\over
N}\exp{[i{\bf k}\cdot({\bf R}_{A^\alpha_{j}}-{\bf R}_{A^\alpha_{k}})]}
\times\exp{\{i[{2\pi\over \phi_0}\int ^{{\bf R}_{A^\alpha_{k}}}_{{\bf
R}_{A^\alpha_{j}}}{\bf A}\cdot d{\bf r}]\}} \cr
&&=t_{4,j}\delta_{j,k+1} \ .
\end{eqnarray}
In this notation,  $t_{4,j}=\exp i\{k_{x}\frac{3}{2}b-k_{y}\frac{a}{2}-\pi{\Phi\over\phi_0}[(j-1)+\frac{1}{2}]\}-\exp i\{k_{x}\frac{3}{2}b+k_{y}\frac{a}{2}+\pi{\Phi\over\phi_0}[(j-1)+\frac{1}{2}]\}$.

\begin{eqnarray}
\langle B^\alpha_{k}|H|B^\alpha_{j}\rangle&&=\sum_{nn}{1\over
N}\exp{[i{\bf k}\cdot({\bf R}_{B^\alpha_{j}}-{\bf R}_{B^\alpha_{k}})]}
\times\exp{\{i[{2\pi\over \phi_0}\int ^{{\bf R}_{B^\alpha_{k}}}_{{\bf
R}_{B^\alpha_{j}}}{\bf A}\cdot d{\bf r}]\}} \cr
&&=t_{5,j}\delta_{j,k+1} \ ,
\end{eqnarray}
and $t_{5,j}=\exp i\{k_{x}\frac{3}{2}b-k_{y}\frac{a}{2}-\pi{\Phi\over\phi_0}[(j-1)+\frac{5}{6}]\}-\exp i\{k_{x}\frac{3}{2}b+k_{y}\frac{a}{2}+\pi{\Phi\over\phi_0}[(j-1)+\frac{5}{6}]\}$. Also,

\begin{eqnarray}
(iii)~~~~\langle A^\alpha_{k}|H|A^\beta_{j}\rangle&&=\sum_{nn}{1\over
N}\exp{[i{\bf k}\cdot({\bf R}_{A^\beta_{j}}-{\bf R}_{A^\alpha_{k}})]}
\times\exp{\{i[{2\pi\over \phi_0}\int ^{{\bf R}_{A^\alpha_{k}}}_{{\bf
R}_{A^\beta_{j}}}{\bf A}\cdot d{\bf r}]\}} \cr
&&=t_{6,j}\delta_{j,k} \ ,
\end{eqnarray}
where $t_{6,j}=\exp i[k_{y}a+2\pi{\Phi\over\phi_0}(j-1)-\frac{\pi}{2}]+\exp i[-k_{y}a-2\pi{\Phi\over\phi_0}(j-1)+\frac{\pi}{2}]$. We also introduce

\begin{eqnarray}
\langle B^\alpha_{k}|H|B^\beta_{j}\rangle&&=\sum_{nn}{1\over
N}\exp{[i{\bf k}\cdot({\bf R}_{B^\beta_{j}}-{\bf R}_{B^\alpha_{k}})]}
\times\exp{\{i[{2\pi\over \phi_0}\int ^{{\bf R}_{B^\alpha_{k}}}_{{\bf
R}_{B^\beta_{j}}}{\bf A}\cdot d{\bf r}]\}} \cr
&&=t_{7,j}\delta_{j,k} \ ,
\end{eqnarray}
with $t_{7,j}=\exp i\{k_{y}a+2\pi{\Phi\over\phi_0}[(j-1)+\frac{1}{3}]-\frac{\pi}{2}\}+\exp i\{-k_{y}a-2\pi{\Phi\over\phi_0}[(j-1)+\frac{1}{3}]+\frac{\pi}{2}$\}.

\begin{eqnarray}
\langle A^\alpha_{k}|H|A^\beta_{j}\rangle&&=\sum_{nn}{1\over
N}\exp{[i{\bf k}\cdot({\bf R}_{A^\beta_{j}}-{\bf R}_{A^\alpha_{k}})]}
\times\exp{\{i[{2\pi\over \phi_0}\int ^{{\bf R}_{A^\alpha_{k}}}_{{\bf
R}_{A^\beta_{j}}}{\bf A}\cdot d{\bf r}]\}} \cr
&&=t_{8,j}\delta_{j,k+1} \ ,
\end{eqnarray}
with $t_{8,j}=\exp i\{k_{x}\frac{3}{2}b+k_{y}\frac{a}{2}+\pi{\Phi\over\phi_0}[(j-1)+\frac{1}{2}]-\frac{\pi}{6}\}+\exp i\{k_{x}\frac{3}{2}b-k_{y}\frac{a}{2}-\pi{\Phi\over\phi_0}[(j-1)+\frac{1}{2}]+\frac{\pi}{6}$\}. Further, we have

\begin{eqnarray}
\langle B^\alpha_{k}|H|B^\beta_{j}\rangle&&=\sum_{nn}{1\over
N}\exp{[i{\bf k}\cdot({\bf R}_{B^\beta_{j}}-{\bf R}_{B^\alpha_{k}})]}
\times\exp{\{i[{2\pi\over \phi_0}\int ^{{\bf R}_{B^\alpha_{k}}}_{{\bf
R}_{B^\beta_{j}}}{\bf A}\cdot d{\bf r}]\}} \cr
&&=t_{9,j}\delta_{j,k+1} \ ,
\end{eqnarray}
with $t_{9,j}=\exp i\{k_{x}\frac{3}{2}b+k_{y}\frac{a}{2}+\pi{\Phi\over\phi_0}[(j-1)+\frac{5}{6}]-\frac{\pi}{6}\}+\exp i\{k_{x}\frac{3}{2}b-k_{y}\frac{a}{2}-\pi{\Phi\over\phi_0}[(j-1)+\frac{5}{6}]+\frac{\pi}{6}$\}.
By diagonalizing our Hamiltonian, the eigenenergy $E^{c,v}$ and the wave function $\Psi^{c,v}$ are derived ($c$ and $v$ refer to the conduction and valence bands, respectively).

\medskip
\par

When a uniform magnetic field $B$ is applied to silicene,
the electronic states are  dispersionless Landau levels (LLs)
whose behavior is governed by the  energy dispersion in the absence
of a magnetic field. Due to the existence of  two valleys
 and spin degrees of freedom,  each LL is eight-fold degenerate  with  successive LL spacing decreasing with
increasing energy. The number of  nodes of an  occupied or unoccupied LL wavefunction
is equal to  the quantum number $n^{c}$ ($n^{v}$ of  each conduction (valence) LL
below or above the chemical potential. We note that the  $n=0$ LL is four-fold
degenerate. Electrons may be excited from valence  LLs to conduction LLs in doped silicene
through electron energy loss spectroscopy (EELS)   or when light  is absorbed, for
example. However, in addition to single-particle excitations  (SPEs) between LLs, there are collective
magnetoplasmon modes whose frequencies are depolarization shifted due to the Coulomb
interaction and are dispersive functions of the transfer momentum $\hbar q$. In our notation,
we label each inter-Landau level excitation  channel  by  ($n^{v}$, $n^{c}$) and the order  of the transition by  $\triangle n=|n^{v}-n^{c}|$.

The dispersion relation for the spectrum of collective plasmon modes may be determined
from  the energy loss  function to be evaluated from    the imaginary part of the inverse
 dielectric function $1/\epsilon(q,\omega)$ where in the random-phase approximation (RPA), we
have $\epsilon(q,\omega)=1-v(q) \Pi(q_\parallel,\omega) $ with
the 2D polarization function   given by \cite{MAG1,MAG2}

\begin{eqnarray}
\label{pol}
 \Pi(q_\parallel,\omega) &=& \frac{g_{s}g_{v}}{(2\pi r_B) ^{2}}
\sum_{n=0}^\infty\sum_{n^\prime=0}^\infty \sum_{s(n),s(n^\prime)}
\frac{f_0(\epsilon_{s(n)n})
-f_0(\epsilon_{s (n^\prime)n^\prime})}
{\hbar\omega-\epsilon_{s(n)n}+\epsilon_{s (n^\prime)n^\prime}+i\gamma}
F_{s(n)s( n^\prime)}(n,n^\prime,q_\parallel)\ ,
\end{eqnarray}
in terms  of the   form factor $F_{s(n)s( n^\prime)}(n,n^\prime,q_\parallel)$     arising from
the overlap of the eigenstates.
The equilibrium Fermi-Dirac distribution  function is
$f(E)=1/[1+exp(E-\mu/k_{B}T)]$, where $k_{B}$ is Boltzmann's
constant. Additionally, $\gamma$ is an energy broadening parameter which may arise
from various dephasing mechanisms, and $\mu$ is the chemical potential whose
temperature dependence may be neglected over the range we investigated.
Also, $v(q)=2\pi e^2/\epsilon_s q$ where
$\epsilon_s=4\pi\varepsilon_0\epsilon_b$ is given in terms of the permeability of free
space and the background dielectric constant $\epsilon_b=2.4$ of silicene. Corresponding to
 each  Landau level  transition  channel, the Re $\epsilon(q,\omega)$ has a symmetric peak
when plotted as a  function of frequency  and a pair of asymmetric peaks. If these peaks occur  where
 $\Im m\ \epsilon(q,\omega)$ vanishes, then  they determine the frequencies of undamped magnetoplasmon
excitations.

\bigskip
\bigskip
\centerline {\textbf {III. RESULTS AND DISCUSSION}}%
\bigskip
\bigskip

The main difference between the LL spectrum of silicene and that of graphene is the splitting of the $n=0$ LL which is due to the significant SOC. Also, a larger Brillouin zone in silicene means that there is \textbf{a higher carrier density per LL} and a wider energy spacing than graphene for a chosen magnetic field. This could make it easier to observe the integer quantum Hall effect at the ambient condition. Another property of silicene is the double-gap energy band structure under an external electric field $E_{z}$, and the transition from topological insulator (TI) to band insulator (BI) when $E_{z}=E_{c}$. Such features are illustrated in Fig. 1 for $E_{z}$-dependent LL spectrum. Based on the node structure of the Landau wavefunctions, the quantum number $n^{c}$ ($n^{v}$) for each conduction (valence) LL could be determined by the number of zeroes; this is the same as the number labeling the $n$th unoccupied (occupied) LL above (below) the Fermi energy $E_{F}$=0 []. (The two localization positions correspond to the two valley structures.) A very small band gap ($\approx$7.9 meV) already exists at $E_{z}=0$. That is, the $n=0$ LL is twice less degenerate than the $n\geq 1$ LLs. A finite $E_{z}$ splits the LLs and lifts the spin and valley degeneracies for the n=0 LLs. This creates two energy gaps with one of them increasing with $E_{z}$ and the other decreasing for $E_{z}<E_{c}$. The lowest gap is closed at $E_{z}=E_{c}$ ($E_{c}\approx\lambda_{SO}/\ell=17\ meV{\AA}^{-1}$), meanwhile the difference between the two energy gaps reaches the maximum. After that, both the two gaps are increased with the increment of $E_{z}$. The process is a signature of band inversion associated with the transition between the TI to BI regimes. The splitting of $n\neq0$ LLs is only obvious at larger $E_{z}'s$ ($E_{z}>1.5$ $E_{c}$).


\medskip
\par

Through the Coulomb interaction, electrons may be excited from the valence  to conduction LLs for the intrinsic condition. A single-particle mode has an excitation energy $\hbar\omega_{ex}=E_{n^{c}}(k+q)-E_{n^{v}}(k)$ according to the conservation laws of energy and momentum. In the discussion below, we use a pair of numbers ($n^{c,v}$,$n^{c,v}$) to label a LL transition channel. Here, $n^{v}$ and $n^{c}$ denote the quantum numbers of the initial and final states, respectively. For example, ($0^{v}$,$1^{c}$) denotes the transition from the highest occupied LL to the second-lowest unoccupied LL, and has the same excitation energy as ($1^{v}$,$0^{c}$) because of the inversion symmetry between the conduction and valence LLs. The transition order is denoted as $\triangle n=|n^{v}-n^{c}|$, which is useful to categorize the SPE channels. The number pair is also used to label a plasmon mode, meaning that the plasmon mode is predominantly dominated by the SPE channel. This is seen in the plasmon frequency that approaches the SPE energy in the large or small limit of $q$, as discussed later in Fig. 4.

\medskip
\par

The SPE spectrum is obtained by calculating  the imaginary part $\epsilon_{2}$ of the dielectric function (Fig. 2). Each prominent peak in $\epsilon_{2}$ represents a major LL transition channel. Their intensity is determined by the wave function overlap between the initial and final states, as given in the Coulomb-matrix elements $|\langle n;\textbf{k}+\textbf{q}|e^{i\textbf{qr}}|m;\textbf{k}\rangle|^{2}$, and therefore strongly depends on $q$. A SPE channel with smaller transition order ($\triangle n=\mid n-m\mid$) and quantum numbers has larger value of the Coulomb-matrix elements for a smaller $q$, and the converse is true for a channel with larger transition order and quantum numbers. This is according to the characteristics of the Hermite polynomials. For $q=1$ (in unit of $10^{5}$ $cm^{-1}$) and Fermi energy $E_F=0$, as shown in Fig. 2(a) by the black curve, the three-lowest frequency peaks are labeled by ($0^{v}$,$1^{c}$), ($1^{c}$,$2^{c}$), and ($2^{c}$,$3^{c}$) from low to high. They represent the interband channels with $\triangle n=1$. The effect of the Fermi level variation is shown in Figs. 2(b)-(e) by the black curves. When $E_{F}$ lies between the n=0 and 1 LLs (Figs. 2(b) and 2(c)), it causes the interband feature of ($0^{v}$,$1^{c}$) to decrease in intensity by a factor of two as half the spectral weight is shifted to a low-energy intraband peak ($0^{c}$,$1^{c}$). When $E_{F}$ is situated between the n=1 and 2 LLs (Fig. 2(d)), the peaks associated with transitions to and from the n=0 LLs disappear due to Pauli Blocking, and the intraband peak $(1^{c},2^{c})$ emerges. The lowest intraband channel $(1^{c},2^{c})$ has a quite lower frequency and a quite stronger intensity than that of the lowest interband channel $(0^{v},1^{c})$ owing to the reduced LL spacings at higher energies and the in-phase Landau wavefunction transition. Peak $(1^{c},2^{c})$ is diminished and replaced by the intraband peak $(2^{c},3^{c})$ as the transitions to the n=1 LL are Pauli blocked ((Fig. 2(e)). With higher $E_{F}$, a similar redistribution of spectral weight is observed.



A finite $E_{z}$ would lift the spin and valley degeneracies of n=0 LLs and enrich the SPE spectrum, as shown by the red curves in Figs. 2(a)-2(e) for $E_{z}=0.5$ $E_{c}$. At $E_{F}=0$ (Fig. 2(a)), the peak $(0^{v},1^{c})$ is split into two: the lowest and the second-lowest interband peaks. The movement of the lowest (second-lowest) interband peak to lower (higher) frequency corresponds to the closing (increasing) band gap. The first feature reaches to the lowest frequency ($\omega_{ex}=E_{n=1}$) when the band gap completely closes at $E_{z}=E_{c}$. For $E_{z}>E_{c}$, both two split peaks move to higher energies due to the reopening of the lowest gap. When $E_{F}$ is above three n=0 levels (Fig. 2(b)), the lowest interband peak redistributes its spectral weight to a lower intraband peak. If $E_{F}$ is further moved to above all n=0 LLs and below the n=1 LLs (Fig. 2(c)), the second-lowest interband peak redistributes its spectral weight to the lowest intraband peak. Therefore, four robust spin- and valley-polarized peaks are obtained. When $E_{F}$ is situated above the n=1 LL (Figs. 2(d)) or higher (Figs. 2(e)), the spin- and valley-polarized peaks could be only observed at higher q's for that the higher transitions from and to the n=0 LLs are allowed.


\medskip
\par

The real parts of dielectric functions are connected with the imaginary parts by the Kramers-Kronig relations (Figs. 3(a)-3(e)). A symmetric peak in $\epsilon_{2}$ corresponds to a pair of asymmetric peaks in $\epsilon_{1}$. The asymmetric peaks could induce zero points. If the zero point exists where $\epsilon_{2}$ vanishes, then an undapmed plasmon could occur at that frequency. In the case of $q=1$ and $E_F$ below the $n=1$ LL (Figs. 3(a)-3(c)), a zero $\epsilon_{1}$ point induced by the channel $(0^{v},1^{c})$ is located where other SPE channels are weak (indicated by the purple arrows). \textbf{Therefore, an undamped plasmon is expected to appear there.} Though the intraband peak ($0^{c}$,$1^{c}$) creates a zero point also (indicated by the blue arrows in Figs. 3(b) and 3(c)), it is located at a finite value of $\epsilon_{2}$ from the adjacent interband channel ($0^{v}$,$1^{c}$). Consequently, only a strongly damped plasmon could occur there. If $E_F$ is above the $n=1$ LL (Fig. 3(d)), a lower zero $\epsilon_{1}$ point is collectively induced by the intraband channels $(1^{c},2^{c})$ and $(1^{c},3^{c})$. The derivative of $\epsilon_{1}$ with respect to frequency around the zero point is relatively small compared to that created by the interband channel ($0^{v},1^{c}$). Therefore, a plasmon with a lower frequency and a stronger intensity is predicted. For an even higher $E_F$ (Fig. 3(e)), more intraband channels involve the zero $\epsilon_{1}$ point. At a high-limit value of $E_F$ , all the low-frequency transition states may collectively contribute to one zero $\epsilon_{1}$ point, which is similar to a classical dielectric form.

\medskip
\par

The energy-loss function, defined as  $\Im m[-1/\epsilon(q,\omega)]$, is useful for understanding the collective excitations and the measured excitation spectra, such as in the inelastic light and electron scattering spectroscopies. Each prominent structure in $\Im m[-1/\epsilon]$ may be viewed as a plasmon excitation with different degrees of Landau damping. For $q=1$ and $E_F=0$ (the black curve in Fig. 4(a)), the lowest and also the strongest peak is located between the SPE energies of $(0^{v},1^{c})$ and $(0^{v},2^{c})$, where corresponds to a zero point in $\epsilon_{1}$ and a quite small value in $\epsilon_{2}$. The second and the third peaks, which are close to the SPE energies of $(1^{v},2^{c})$ and $(2^{v},3^{c})$, respectively, are relatively weak due to significant Landau damping.

When $E_{F}$ lies between the n=0 and 1 LLs (Figs. 4(b) and 4(c)), the spectral weight of the first plasmon peak is shifted to a low-energy intraband plasmon. If $E_{F}$ is between the $n=1$ and 2 levels (Fig. 2(d)), the threshold peak is replaced by a lower intraband plasmon which is contributed by channels $(1^{c},2^{c})$ and $(1^{c},3^{c})$. The intraband plasmon involves more low free-charge carriers and has the stronger intensity than any interband features. For an even higher $E_{F}$, the intraband plasmon grows in intensity obviously, a result of the increased (decreased) number of intraband (interband) transition channels plotted in the blue (red) dashed lines in Fig. 4(e). A finite $E_{z}$ ($<E_{c}$) would lower the threshold-excitation frequency due to the splitting of plasmon peaks ($0^{v},1^{c}$) and ($0^{c},1^{c}$), as shown in Figs. 4(a)-4(c) by the red curves. The newly created peaks suffer a quite strong Landau damping since the splitting energies between the n=0 LLs are quite small. The splitting energies may be improved by a stronger $E_{z}$, and then the low plasmon peaks are enhanced. This is illustrated by the blue curves in Figs. 4(a)-4(c) for $E_{z}=2$ $E_{c}$. When $E_{F}$ is situated above the n=1 LL or higher (Figs. 4(d) and 4(e)), the effect of the E-field is only evident at higher q's. This is seen in the plasmon dispersion later.

\medskip
\par

The magnetoplasmon spectrum possesses an intriguing dependence on $q$, as shown in Fig. 5. Each plasmon branch is strongly confined in between the energies of two close SPE channels (illustrated by the two white-dashed lines for the lowest plasmon branch) and exists only in a limited $q$-range. In both the short and long wavelength limits, plasmons are overly damped and have frequencies close to the SPE energies. Therefore, a characteristic behavior of the magnetoplasmons is that in the long wavelength limit their group velocity is positive and the plasmon intensity is increased as $q$ is increased. The group velocity becomes zero at a critical momentum $q_{B}$ where the length scale for density fluctuations is comparable to the cyclotron radius. As for $q>q_{B}$, the group velocity becomes negative and the plasmon intensity is decreased by further increasing $q$. The value of $q_{B}$ is increased when the magnetic field $B$ is increased. For a plasmon-excitation channel, the larger the transition order $\triangle n$, a larger rate of increase in $q_{B}$ as a function of  $B$ is obtained [ACS NANO 5,1026 (2011)]. The peculiar dependence of $q_{B}$ on $B$ may cause a rich $B$-dependent plasmon spectrum, as demonstrated in later plots.

\medskip
\par

The dispersion relation of the intraband magnetoplasmons is quite different from the interband ones'. When the $n=1$ LL is occupied (Fig. 5(b)), the lowest interband plasmon branch $(0^{v},1^{c})$ no longer exists and is replaced by a combination of three intraband modes: $(1^{c},2^{c})$, $(1^{c},3^{c})$ and $(1^{c},4^{c})$. The three intraband channels form a continuous branch which exhibits a longer range of positive slope (group velocity) and higher intensity. Besides, the disappearance of the interband plasmon $(1^{v},1^{c})$ helps to enhance the other interband modes, such as $(0^{v},3^{c})$ and $(0^{v},4^{c})$. Although these interband plasmons are close to each other, they disperse independently, because they experience quite a strong restoring force which comes from the magnetic field. The intraband magnetoplasmon would extend to the higher frequency for a higher Fermi level (Fig. 5(c)) owning to the increased number of intraband channels. Meanwhile, the gap between the intraband plasmon and the lowest interband plasmon is reduced. A finite $E_{z}$ would induce extra branches mainly out of the splitting of $n=0$ LLs, as shown in Fig. 5(d) for $E_{F}=70$ meV and $E_{z}=2$ $E_{c}$. Those newly created modes (marked by the purple arrows) are weakly dispersive. The locations and the number of the sub-branches vary with the Fermi level. They could be used as a fine tuning of the magnetoplasmon spectrum, like the threshold-excitation frequency, the energy spacings and the density of excitation channels.


\medskip
\par

The plasmon spectra at various $q$'s and $E_F$'s have different dependencies on the magnetic field. For $E_F=0$ and a small q=1, the plasmon modes are dominated by the SPE channels with $\triangle n=1$ according to the characteristics of Hermite polynomial functions. The frequencies and intensities of these modes grow monotonically with $B$ (Fig. 6(a)), which corresponds to an enhanced carrier density per LL and enlarged LL spacings. When $q$ is increased to 5 (Fig. 6(b)), the plasmon modes with $\triangle n=2$ become more intense. Such channels have the larger $q_{B}'s$ than those of $\triangle n=1$ and hardly exist at the long wavelength limit ($q\to 0$). This is because the plasmon intensity is reduced when $q$ is away from $q_{B}$. The critical momentum $q_{B}$ grows with increased $B$ on the account of the increased ratio of the wavelength to the magnetic length. Therefore, if $q$ is set to a value less than $q_{B}$ for a plasmon mode, an increasing $B$ in effect causes $q$ to deviate from $q_{B}$ that weakens the plasmon further. This is seen in the branches of $(2,0)$ and $(4,0)$ in Fig. 6(b). On the other hand, if $q$ is set to a value larger than $q_{B}$, an increasing $B$ will move $q_{B}$ closer to the fixed value of $q$. All plasmon modes in such the condition are intensified consistently by an increasing $B$, as shown in Fig. 6(c) for $q=10$.


\medskip
\par

In the condition of that $n=1$ LL is occupied (Fig. 7), the plasmon spectrum experiences an abrupt change in the intensities, frequencies, and bandwidths at the critical field strength $B_{c}\approx5.5$ T. This is because that above the $B_{c}$ all electrons may be accommodated in the $n=0$ LLs.  For $q=1$ (Fig. 7 (a)), the lowest dominant transition channel is changed from the intraband $(1^{c},2^{c})$ to the interband $(0^{v},1^{c})$ at the $B_{c}$. In contrast, the other interband channels are increased in intensities and frequencies continuously. At $q=5$ (Fig. 7(b)), the obvious change in the plasmon spectrum after the cross of $B_{c}$ is the discontinuous transformation from the lowest intraband channel $(1^{c},4^{c})$ to the lowest interband $(0^{v},1^{c})$ and the appearance of the interband branch $(1^{v},1^{c})$.  The distinct $B$-dependent behaviors between the intraband and interband plasmon modes is a vivid feature to distinguish the two kinds of transition channels. At $q=10$, there are more channels from and to the n=1 LL , like (5,1) and (8,1), as shown in Fig. 7(c). These channels disappear at the $B_{c}$ and create more discontinuous features. The result demonstrates that the momentum transfer is an important factor to perform a tunable B-dependent plasmon spectrum.

If both the $n=1$ and $n=2$ LLs are occupied, there are two $B_{c}$'s through which $n_{F}$ (the quantum number of the highest occupied LL) is decreased to $n_{F}-1$, as shown in Fig. 8. The first critical magnetic field is $B_{c1} \sim5.5$ T (the same with that in Fig. 7), and the second is $B_{c2} \sim11$ T. For q=1 (Fig. 8(a)), the lowest intraband channel $(2^{c},3^{c})$ is changed to intraband $(1^{c},2^{c})$ at $B_{c1}$, and then the intraband channel ($1^{c},2^{c}$) is further changed to interband ($0^{v},1^{c}$) at $B_{c2}$. The conditions of various q's are displayed in Figs. 8(b) and 8(c). Generally speaking, the plasmon spectrum are distinct in the three field-strength ranges: $0<B<B_{c1}$, $B_{c1}<B<B_{c2}$, and $B_{c2}<B$. The three regions show the features of the LL spectrum with the highest occupied quantum number: n=2, n=1, and n=0, respectively. The discontinuous plasmon behaviors at $B_{c}'s$ could be verified experimentally. The novel feature with the strong dependence on wave vector as well as the excitation channels may be useful for the design of magneto-plasmonic components for varied applications.

\bigskip
\bigskip
\centerline {\textbf {IV. SUMMARY AND CONCLUSIONS}}%
\bigskip
\bigskip

In summary, the magnetoplasmon behaviors in monolayer silicece are studied by the novel GTBM. GTBM simultaneously includes the atomic interaction, the spin-orbit coupling, the Coulomb interaction, and the effect of external fields. The buckling structure and the significant SOI warrant an electrically tunable energy gap and spin- and valley- polarized LLs in a magnetic field. This reflects in the energy spacing and the number of magnetoplasmon peaks. The extra plasmon branches induced by the E-field have weak dependence on the momentum transfer, which means that the buckling structure and the magnetic field combine to localize them. The heightened Fermi level is shown to redistribute spectral weight from discrete interband transitions to a strong low-energy intraband plasmon. The two plasmon modes differ in the dependence on the momentum and the B-field strength. In particular, discontinuous changes in the plasmon features are found when field strength is equal to $NB_{c}$. At the crossing of each $B_{c}$, the quantum number of the highest occupied LL is changed from $n_{F}$ to $n_{F}-1$. We demonstrate that the characteristic B-dependent plasmon behaviors are strongly determined by the momentum transfer. This is due to the different $q_{B}'s$ of each plasmon branch (plasmon at $q_{B}$ has zero group velocity). The methodology is suitable to treat a large variety of forms of external fields, including uniform and nonuniform electric and magnetic fields. Also, it can be extended to other nanomaterials at a wide range of chemical potentials. Therefore, this work suggests a guide in searching the materials for better nanoplasmonic applications.

\emph{In summary, we have calculated the charge  carrier plasmon dispersion relation in silicene under conditions of variable perpendicular  electric and magnetic fields as well as the adjustment of free carrier density is adjusted by doping the sample.  The non-dispersive Landau level energy bands may be characterized as being in two spin-up and spin-down  subgroups due to the  presence of spin-orbit coupling. A major effect arising from the applied electric field on the LLs is to lift the degeneracy of the up and down spin states for intrinsic silicene.
On the other hand, in the absence of an electric field, doping would create more separated plasmon
branches, which involve the two separated $n=0$ LLs. Consequently, the LLs may be split and
inter-LL transitions may lead to a novel quasiparticle excitation spectrum which  is not possible
to achieve without these three contributing factors being applied simultaneously. Most notably, we
observed the appearance of a group of dispersionless plasmon excitations which means that the
buckling geometry  and magnetic field combine to localize them.}

\medskip
\par

\emph{Our analysis has demonstrated that the plasmon dispersion of 2D systems can be even more easily tuned than
their 3D counterparts when carrier doping, electric and magnetic fields together act.
The negative dispersion in the plasmon spectrum  can be switched to positive by
doping with electrons or holes.  Additionally, the plasmon nature  is determined
by an interplay between the Coulomb interaction and the interband LL transitions, which gives
rise to a small bandwidth corresponding to localized plasmons at high doping as shown in
Fig.\ 5. Thus, we have managed to tune the plasmon character  microscopically
through the charge response to a strong  local field in silicene.
Therefore,  doped silicene in perpendicular electromagnetic fields
seem  promising for further investigation experimentally.}

\bigskip

\bigskip

\centerline {\textbf {ACKNOWLEDGMENT}}%

\bigskip

\bigskip

\noindent \textit{Acknowledgments.} This work was supported by the NSC of Taiwan, under Grant No. NSC 102-2112-M-006-007-MY3.

\newpage

\par\noindent ~~~~$^\star$e-mail address: yarst5@gmail.com

\par\noindent ~~~~$^\dag$e-mail address: ggumbs@hunter.cuny.edu

\par\noindent ~~~~$^\dag$$^\dag$e-mail address: mflin@mail.ncku.edu.tw

\newpage

\begin{figure}
\includegraphics[width=1.0\textwidth]{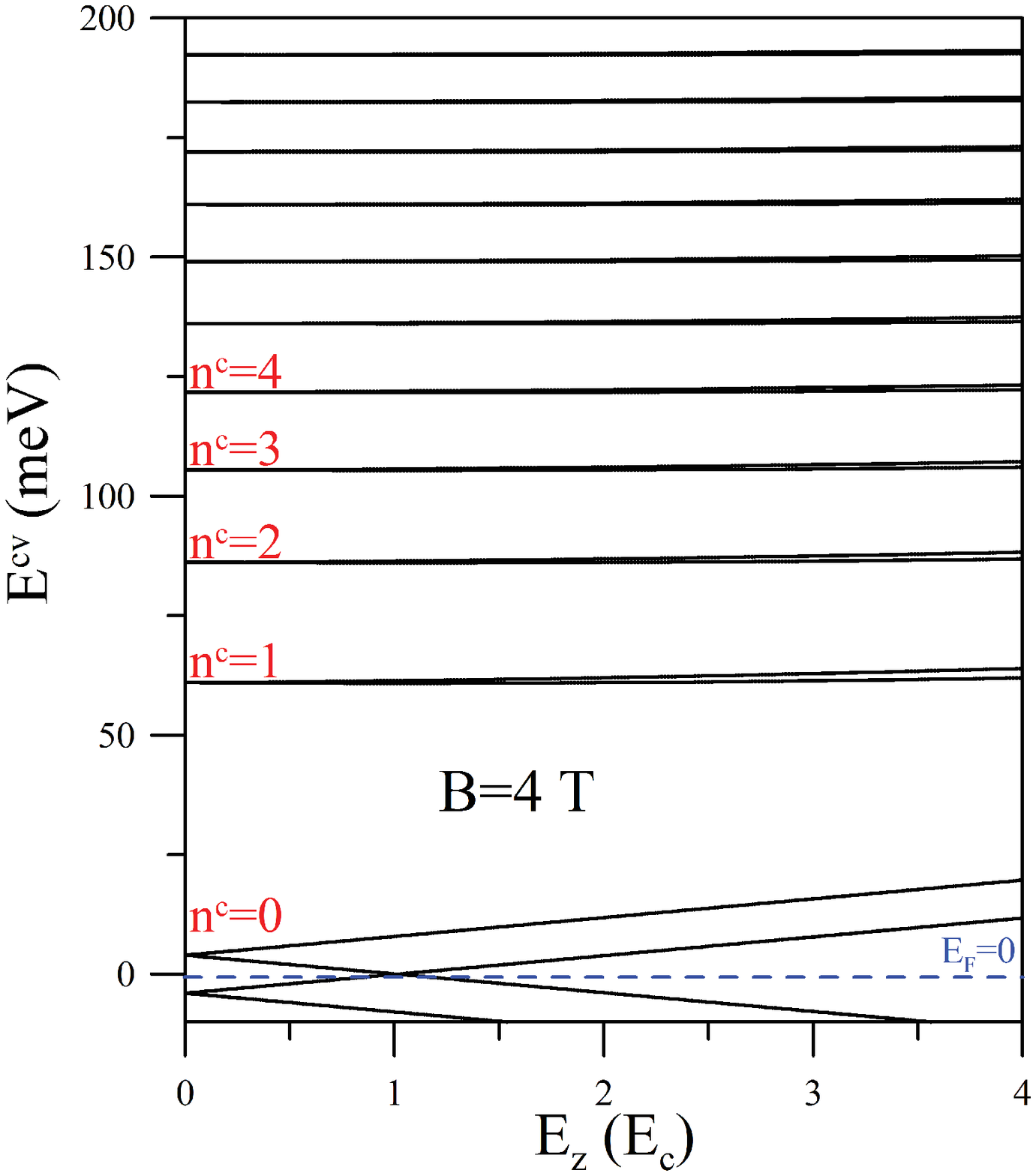}
\caption{The Landau level energies as a function of the $E_z$
electric field in units of the critical field $E_c$ \  ($\sim 17$ meV $\AA^{-1}$).}
\end{figure}

\begin{figure}
\centering
\includegraphics[width=0.5\textwidth]{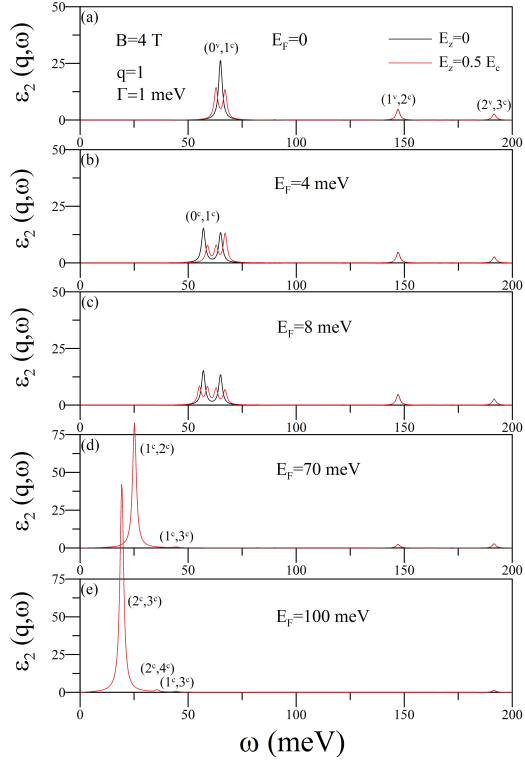}
\caption{(Color online) The imaginary part of the dielectric
functions  $\epsilon(q,\omega)$ for a chosen wave vector transfer $q=1$ and different
Fermi energies $E_F$. The black and red curves correspond to $E_z=0$ and $E_z=0.5\  E_c$, respectively.}
\end{figure}

\begin{figure}[H]
\centering
\includegraphics[width=0.75\textwidth]{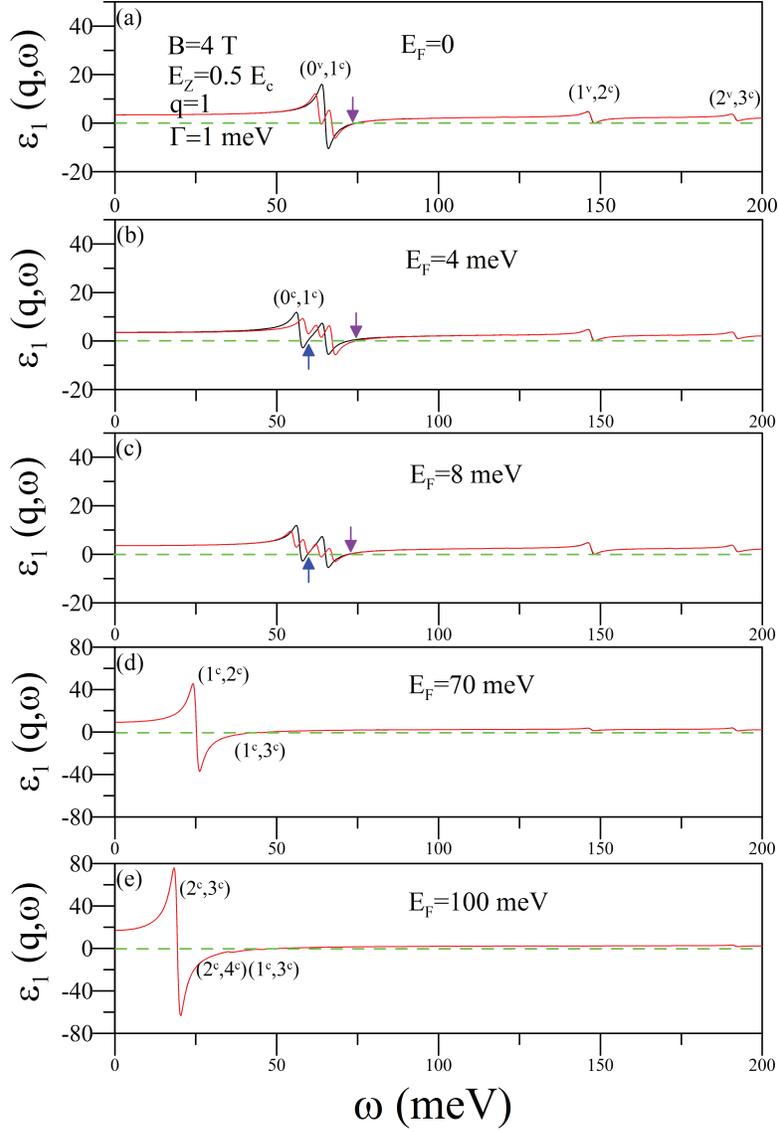}
\caption{The real part of the dielectric functions $\epsilon(q,\omega)$ for a chosen wave vector transfer $q/k_F=1$ and different
Fermi energies $E_F$. The black and red curves correspond to $E_z=0$ and $E_z=0.5\  E_c$, respectively.}
\end{figure}

\begin{figure}[H]
\centering
\includegraphics[width=0.75\textwidth]{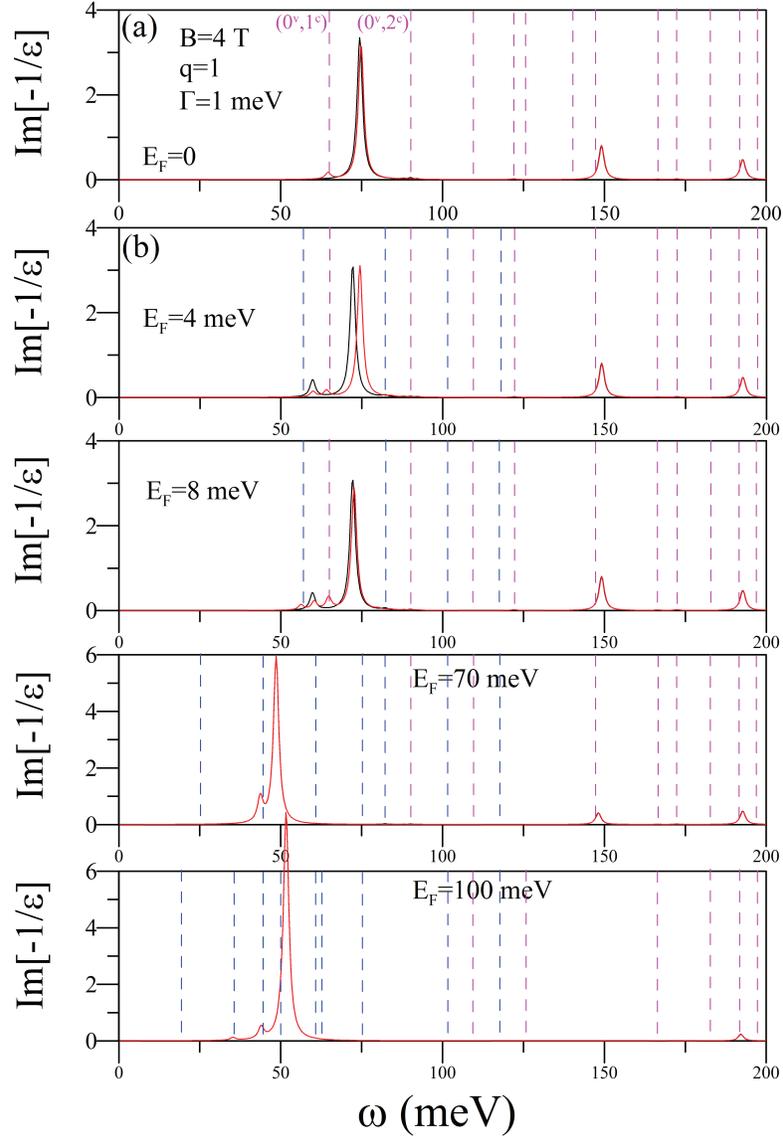}
\caption{The energy-loss function for chosen $q=1$ and different Fermi energies.}
\end{figure}

\begin{figure}[H]
\centering
\includegraphics[width=1.0\textwidth]{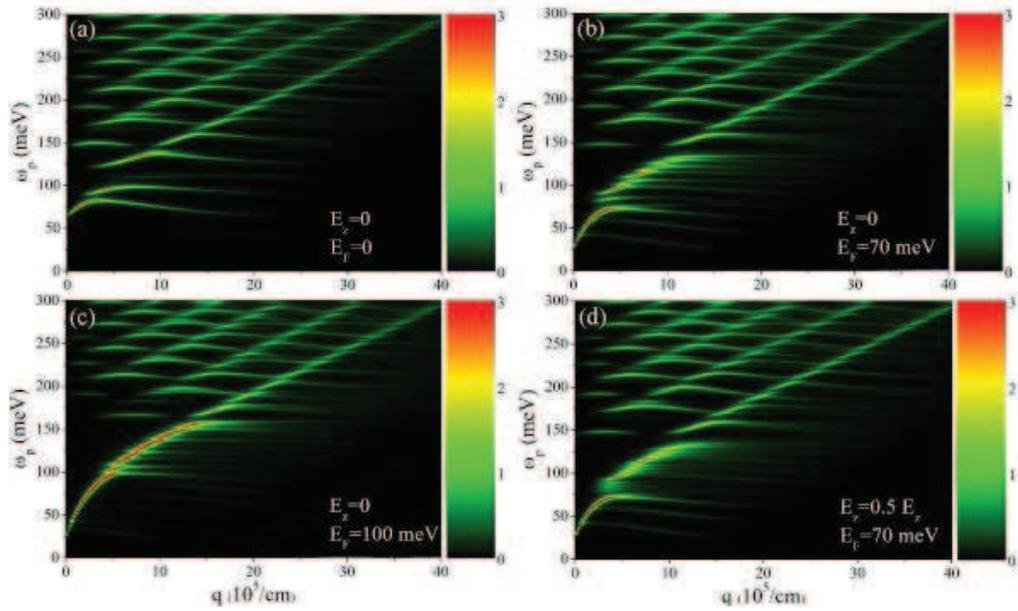}
\caption{(Color online) Density plot of frequency $\omega$ versus wave number $q$ for various
$E_F$. The color scale represents the intensity of the energy loss function
for the excitation of undamped plasmon modes.}
\end{figure}

\begin{figure}
\centering
\includegraphics[width=1.0\textwidth]{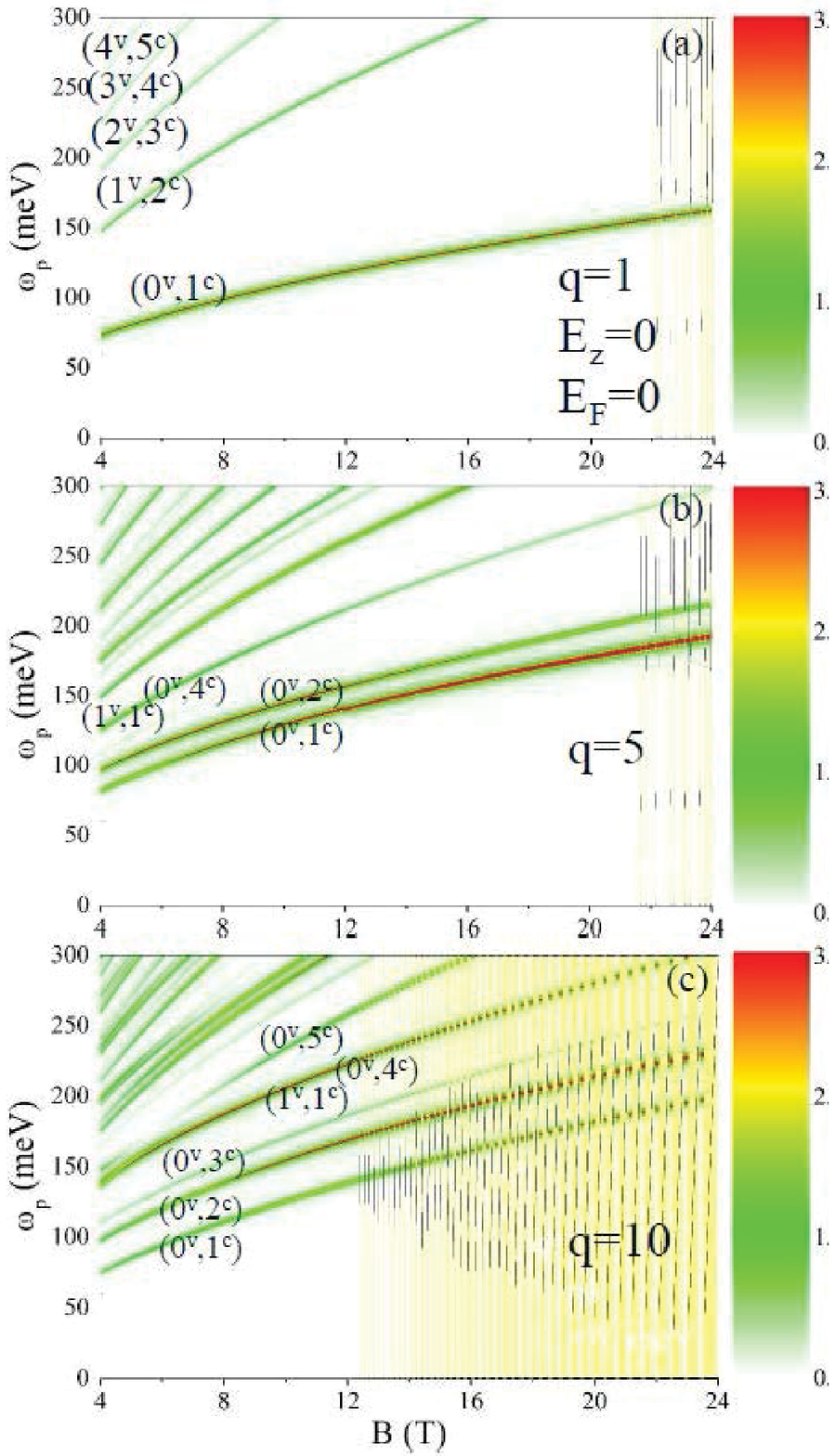}
\caption{(Color online) Density plot of $\omega$ versus $B$ for $E_F=0$  meV.}
\end{figure}

\begin{figure}[H]
\centering
\includegraphics[width=1.0\textwidth]{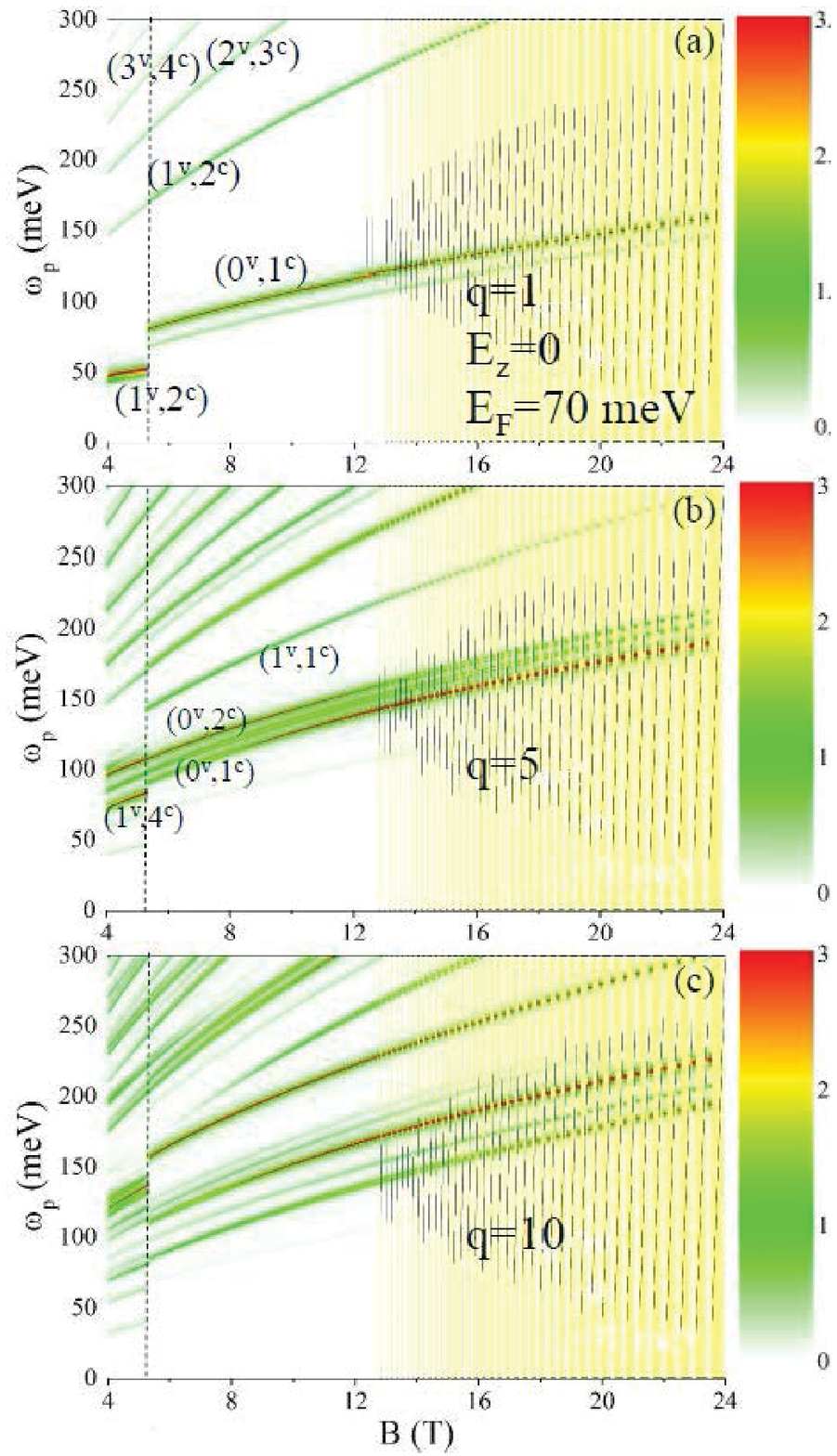}
\caption{(Color online) Density plot of $\omega$ versus $B$ for $E_F=70$  meV.}
\end{figure}

\begin{figure}[H]
\centering
\includegraphics[width=1.0\textwidth]{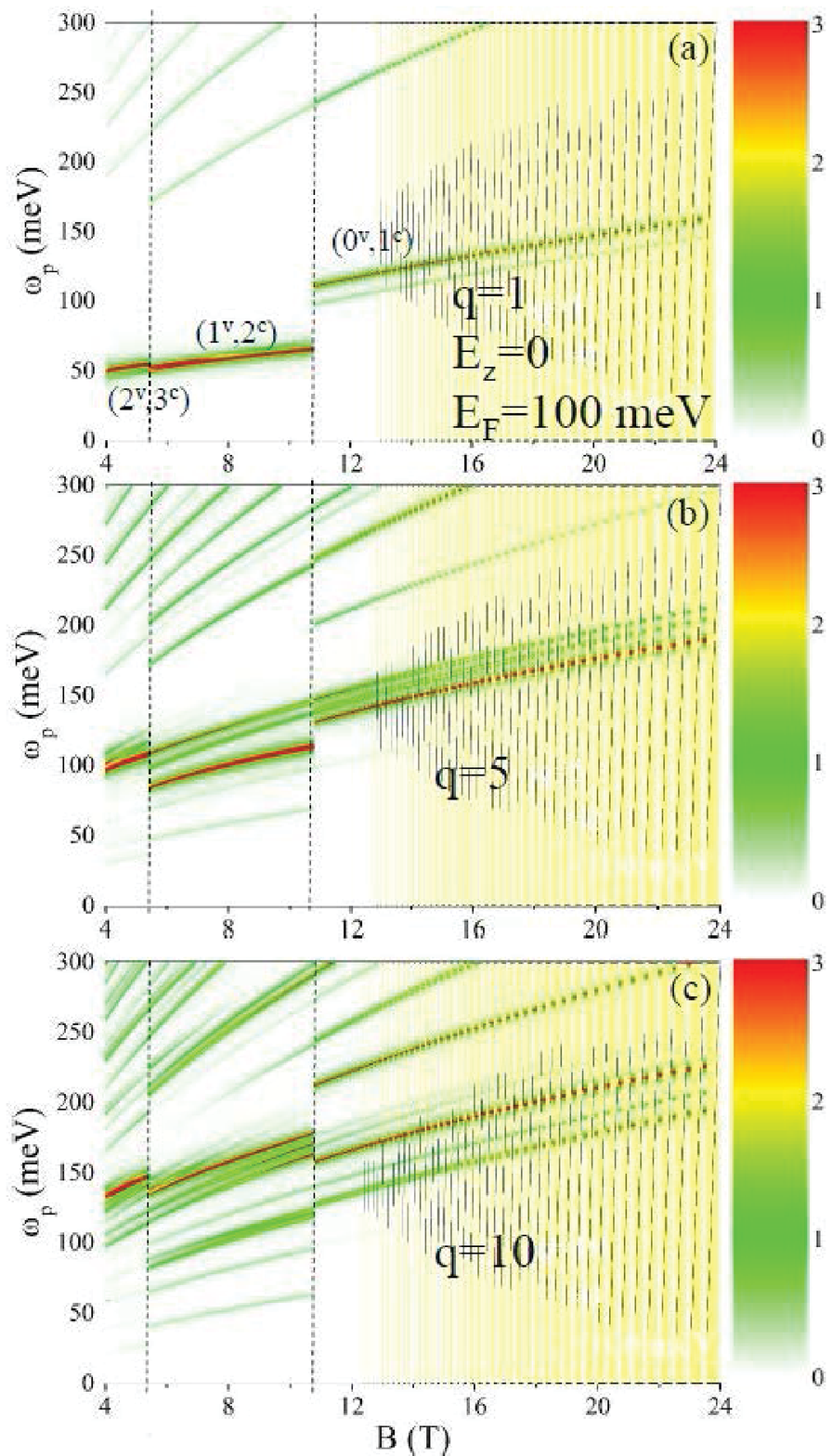}
\caption{(Color online) Density plot of $\omega$ versus $B$ for $E_F=100$  meV.}
\end{figure}

\vskip0.5 truecm

\end{document}